\documentclass[showpacs,pra]{revtex4}
\usepackage{graphicx}
\usepackage{dcolumn}
\usepackage{bm}
\begin{document}
\title{Propagation of short pulses through a Bose-Einstein condensate}

\author{Devrim Tarhan$^{1,2,3}$ , Se\c{c}kin \c{S}efi$^{1}$ and
\"{O}zg\"{u}r E. M\"{u}stecapl\i{}o\~{g}lu$^{2}$}

\address{$^1$Department of Physics, Faculty of Sciences and
Letters,\\
Istanbul Technical University, Maslak 34469, Istanbul, Turkey \\
$^2$Department of Physics, Ko\c{c} University Rumelifeneri yolu,
Sar\i{}yer, Istanbul, 34450, Turkey \\
$^3$Department of Physics, Harran University , Osmanbey
Yerle\c{s}kesi, \c{S}anl\i{}urfa, Turkey}

\email{dtarhan@ku.edu.tr}

\begin{abstract}
We study propagation of short laser pulses in a Bose-Einstein
condensate taking into account dispersive effects under the
conditions for electromagnetically induced transparency. We
calculate dispersion coefficients using typical experimental
parameters of slow-light schemes in condensates. By numerically
propagating the laser pulse, and referring to theoretical
estimations, we determine the conditions for which dispersion
starts to introduce distortions on the pulse shape.
\end{abstract}
 \maketitle \narrowtext
\section{Introduction}
Since the first demonstration of ultra-slow light propagation
through a Bose-Einstein condensate (BEC)\cite{Hau}, a major
application of slow light has been argued to be an optical
memory\cite{Fleischhauer}. Information storage capacity of such an
optical memory based upon slowing laser pulses down to subluminal
speeds in BECs would strongly depend on the temporal width of the
pulses. As the temporal width of the pulses gets smaller,
condensate could contain more pulses in a given time interval.
Unfortunately, one cannot hope for using shortest pulses (such as
femtosecond or even attosecond) available to maximize information
capacity. A challenging issue to use such a short pulse in
condensates is the fact that slow-light scheme in BEC is based
upon electromagnetically induced transparency
(EIT)\cite{Harris1,Boller} which occurs for not so broad optical
frequency window. On the other hand, temporally short pulses have
large frequency widths. Present experimental set ups use optical
pulses of temporal widths in the order of microseconds. We note
that EIT conditions can be reasonably well satisfied for pulses of
widths $\sim 10^{-7}$ which would enhance the information capacity
of the condensate ten times more.

Another challenging issue in using short pulses in BECs is that
their susceptibility to dispersion which may distort the pulse
shape. We note that even though dispersive effects have been found
small experimentally for microsecond pulses, there is no guarantee
that role of dispersion would be small when we make the pulse ten
times shorter. As the earlier theoretical models ignore the
dispersion completely\cite{Mustecaplioglu1,Mustecaplioglu2}, it is
necessary to develop a more general theory of slow light
propagation taking into account the higher order dispersive
properties of the BEC. In the present paper we develop such a
theory, generalizing theory of dispersive  EIT in a thermal
uniform gas\cite{Harris}. Using our general theory, we have found
that dispersive effects are insignificant for microsecond pulses
used in slow-light experiments. However, temporal width of
microsecond is the critical width below which dispersion starts to
distort the pulse shape. We show that at a width of
$10^{-7}\,\mathrm{seconds}$, dispersion leads to a small temporal
broadening of the pulse.

It may be noted here that study of dispersive effects in
slow--short-light propagation may lead to alternative applications
such as determination of condensate temperature through
measurement of broadening, pulse shape engineering, and frequency
band narrowing. While we treat the light classically, novel
effects have been predicted for slow-light in quantum regime, in
particular a new nonlinearity regime for quantum
light\cite{Lukin}. Our results may contribute to such studies as
well. Furthermore, the theory in the present paper can easily be
extended to situations in which BEC can have an enhanced nonlinear
optical response of BEC that might be utilized for dispersion
compensation.

The paper is organized as follows. In Sect. 2, Electric
susceptibility of an interacting BEC under EIT conditions is
derived. In Sect. 3, wave equation including the high order
dispersive terms for the EIT medium is derived. We present and
discuss our numerical results in section 4. Finally, we conclude
in section 5.
\section{EIT Electric susceptibility for an interacting BEC}
EIT is a technique for making cancellation of induced absorption
to weak probe field tuned in resonance to an atomic transition by
applying a strong resonant electromagnetic field to couple
coherently another atomic transition. Under EIT refractively thick
gaseous medium becomes transparent to the probe beam. In order to
formulate the effect, we consider a three-level atom, where the
atomic states are denoted by $|1\rangle$ , $|2\rangle$ and
$|3\rangle$, $\Lambda$-type atom interacting with two lasers.
Interaction of electromagnetic field with such an atom, under
electric dipole approximation, can be described by a Hamiltonian
of the form $H = H_{0} + H_{1}$, where
\begin {eqnarray}
H_{0} &=& \hbar \omega_{1}|1> <1| +\hbar \omega_{2}|2> <2| +
\hbar \omega_{3}|3> <3|, \nonumber\\
H_{1} &=& -e \,\vec{r} \cdot \vec{E}(t).
\end {eqnarray}
Here, $H_{0}$ and $H_{1}$ are unperturbed and interaction parts of
the Hamiltonian, respectively. Probe laser, with angular frequency
$\omega_p$, is tuned to transition $|1>$ to $|3>$ while coupling
laser, with angular frequency $\omega_c$, is tuned to transition
$|2>$ to $|3>$. We have assumed that only $|3>$ $\rightarrow$
$|1>$ and $|3>$ $\rightarrow$ $|2>$ transitions are dipole
allowed. The interaction part of the hamiltonian for the (EIT)
system can be calculated by evaluating
\begin {eqnarray}
H_{1}= -(|1\rangle \langle1| + |2\rangle \langle2| + |3\rangle
\langle3|) e \, \vec{r}\cdot \vec{E}(t) (|1\rangle \langle1| +
|2\rangle \langle2| + |3\rangle \langle3|).
\end {eqnarray}
Denoting dipole matrix elements $\mu_{31} = \mu_{13}^* = e
\langle3|r|1\rangle$ , $\mu_{32} = \mu_{23}^* = e
\langle3|r|2\rangle$ , $\mu_{12} = \mu_{21}^* = e
\langle1|r|2\rangle$ and introducing slowly varying amplitudes for
the electric fields $E_{p}(t)=(\varepsilon/2) \exp(-i \omega t)$
and $E_{c}(t)= (\varepsilon /2) \exp(-i \omega_{c} t)$, Rabi
frequency for the coupling laser can be written as
$\Omega_{c}=\mu_{32}\,\varepsilon/ \hbar$. Interaction part of the
hamiltonian becomes
\begin {eqnarray}
\nonumber H_{1} &=& -\frac{\hbar}{2} \{
\frac{\mu_{31}\,\varepsilon}{\hbar} \exp(-i \omega t) |3\rangle
\langle1|
\\\nonumber &+& \frac{\mu_{13}\,\varepsilon}{\hbar}\exp(i \omega t)
|1\rangle \langle3|
\\ &+& \Omega_{c} \exp(-i \omega_{c} t) |3\rangle \langle2|
+ \Omega_{c} \exp(i \omega_{c} t) |2\rangle \langle3| \}
\end {eqnarray}

Including the decay terms phenomenologically, to take into account
the finite life time of the atomic levels, to the Liouville
Equation $i\hbar\dot{\rho} = [ H , \rho ]$, density matrix
equation become\cite{scully}
\begin {equation}
\label{genellio} \dot{\rho} = -\frac{i}{\hbar} [ H , \rho ] -
\frac{1}{2} \{\Gamma , \rho\}
\end {equation}
where $\Gamma$  is the relaxation matrix,which is defined by the
equation $< n \,| \,\Gamma \, |\, m > = \, \gamma_{n}\,
\delta_{nm}$ . At the same time the relation between $\Gamma$ and
$\rho$ is given by $\{ \Gamma , \rho\} = \Gamma \rho + \rho \Gamma
$. In terms of the matrix elements, explicit equation is given by
(\ref{genellio}) \cite{scully}
\begin {eqnarray}
 \label{genelij}\dot{ \rho_{\mathrm{ij}}} = -\frac{i}{\hbar}
\sum_{\mathrm{k}} (H_{\mathrm{ik}} \rho_{\mathrm{kj}} -
\rho_{\mathrm{ik}}H_{\mathrm{kj}})-\frac{1}{2} \sum_{\mathrm{k}}
(\Gamma_{\mathrm{ik}} \rho_{\mathrm{kj}} + \rho_{\mathrm{ik}}
\Gamma_{\mathrm{kj}}).
\end {eqnarray}

The equation of motion for the density matrix elements $\rho_{31}$
and  $\rho_{21}$ are given by
\begin {eqnarray}
\label{densitymatrixelements} \dot{\rho_{31}} &=& - (i \omega_{31}
+ \frac{\Gamma_3}{2}) \rho_{31} \nonumber \\&-&\frac{i}{2}
\frac{\mu_{31} \varepsilon}{\hbar} \exp(-i \omega t)
(\rho_{33}-\rho_{11}) + \frac{i}{2} \Omega_{c} \exp(-i \omega_{c}
t) \rho_{21}
\\ \dot{\rho_{21}} &=&- (i \omega_{21} + \frac{\Gamma_2}{2})
 \rho_{21} \nonumber\\&-& \frac{i}{2} \frac{\mu_{31} \varepsilon}{\hbar} \exp(-i
\omega t) \rho_{23} + \frac{i}{2} \Omega_{c} \exp(i \omega_{c} t)
\rho_{31}
\end {eqnarray}

The levels $|3>$ $\rightarrow$ $|1>$ are coupled by a probe field
of amplitude $\varepsilon$ and frequency $\omega$ , whose
dispersion and absorption we are interested in. The dispersion and
absorption can be computed by $\rho_{31}$. All atoms are initially
in the ground level $|1>$,  $\rho_{11}^{(0)} = 1,\rho_{22}^{(0)} =
\rho_{33}^{(0)} = \rho_{31}^{(0)} = 0.$ In order to solve equation
(\ref {densitymatrixelements}), we can write the equations in the
matrix form \cite{scully},$\dot{R}=-MR+A$. where $R$, $M$ and $A$
are matrix elements. As a result R is given by $R=M^{-1}A$.

We found for the density matrix element $\rho_{31}$,
\begin {equation}
\rho_{31}=\frac{i \mu_{31} \varepsilon \exp(-i \omega t) (i \Delta
+ \Gamma_2/2)}{2 \hbar [(\Gamma_2/2 + i \Delta)(\Gamma_3/2 + i
\Delta) + \frac{\Omega_c^2}{4}]}
\end {equation}

Finally, we determine the electric susceptibility $\chi$ of BEC
consisting of such three-level atoms in EIT scheme. If we use the
relationship between electric field, atomic polarizability and
(macroscopic) polarization\cite{scully}, we find
\begin {equation}
\label{chieit}\chi = \frac{\rho|\mu_{31}|^2}{\epsilon_0 \hbar}
\frac{i(i \Delta + \Gamma_2/2)}{[(\Gamma_2/2 + i
\Delta)(\Gamma_3/2 + i \Delta) + \frac{\Omega_c^2}{4}]},
\end {equation}
where $\rho=\rho(\vec{r})$ stands for the inhomogeneous
concentration of atoms in an interacting BEC. At a given
temperature density profile of an interacting BEC is composed of
two components\cite{naraschewski}. Density of the condensate part
is evaluated by using Thomas-Fermi approximation, while the
density of the thermal component is evaluated semi-classically in
a harmonic trap\cite {Bagnato}. The total ground state density is
given by
\begin {equation}\label {positiondependentro}
\rho(\vec{r}) = \frac{\mu-V(r)}{U_{0}} \Theta(\mu-V(r))
\Theta(T_c-T) + \frac{g_{3/2} (z e^{-\beta V})}{\lambda_T^3}.
\end {equation}
where $U_0=4 \pi \hbar^2 a_{s}/m$, $m$ is the atomic mass and
$a_s$ is the s-wave scattering length, $\Theta(.)$ is the
Heaviside step function, $g_{n}(x)=\Sigma_{j} x^j/j^n$,
$\lambda_T$ is the thermal de Br\"{o}glie wavelength,
$\beta=\frac{1}{k_{B}T}$, and $T_C$ is the critical temperature.
We assume an external trapping potential in the form
$V(\vec{r})=(1/2) m (\omega_r^2 r^2+\omega_z^2 z^2)$ with
$\omega_r$ the radial trap frequency and $\omega_z$ the angular
frequency in the z direction. $\mu$ is the chemical potential.
\section{Wave equation in dispersive media}
In a dispersive media the relation between displacement current
vector and electric field is given by
\begin {equation}
\label{displacement} D = \epsilon(\omega) E
\end {equation}
where $\epsilon(\omega) = \epsilon_{0} [1+ \chi(\omega)]$.
Polarization is related to the electric field through electric
susceptibility via $P = \epsilon_{0} \chi(\omega) E$.
Susceptibility $\chi(\omega)$ is the Fourier transformation of the
response function (Green function) of the medium $\chi(t)$,
 $\chi(\omega) = \int_{-\infty}^{\infty}
\chi(t)\exp(-i \omega t)dt$. Dispersive media are characterized by
frequency dependence of susceptibility. Expanding the
susceptibility of the dressed atom to second order about a central
frequency $\omega_0$, and with
$P(\omega-\omega_0)\equiv\epsilon_{0}\chi(\omega-\omega_0)E(\omega-\omega_0)$,
yields\cite{Harris}.
\begin {eqnarray} \label{chi}
\chi(\omega-\omega_{0}) &=& \chi(\omega_{0})
+\frac{\partial\chi}{\partial\omega}|_{\omega_{0}}(\omega-\omega_{0})+
\frac{1}{2}\frac{\partial^{2}\chi}{\partial^{2}\omega}|_{\omega_{0}}(\omega-\omega_{0})^2,
\end {eqnarray}
which leads to the polarization\cite{Harris}
 \begin{eqnarray}\label{dispersifpolarization}
 P(t)  &=& \epsilon_{0}\chi(\omega_{0})E(t)
-i\epsilon_{0}\frac{\partial\chi}{\partial\omega}|_{\omega_{0}}
\frac{\partial E}{\partial t} -
\frac{\epsilon_{0}}{2}\frac{\partial^{2}\chi}{\partial^{2}\omega}|_{\omega_{0}}
\frac{\partial^2 E}{\partial^2 t}.
\end{eqnarray}

Maxwell equations in the medium are given by
\begin {eqnarray}
\mathbf{\nabla} \cdot \mathbf{B} &=& 0 \quad
\mathbf{\nabla} \cdot \mathbf{D} = \rho_{free} \\
\mathbf{\nabla} \times \mathbf{E} &=& - \frac{\partial
\mathbf{B}}{\partial t} \quad \mathbf{\nabla} \times \mathbf{H} =
\mathbf{J} +  \frac{\partial \mathbf{D}}{\partial t}.
\end {eqnarray}
In addition, we have the standard constitutive relations
$\mathbf{D} = \epsilon \mathbf{E} + \mathbf{P}$, $\mathbf{B} =
\mu_0 \mathbf{H}$, $\mathbf{J}=\sigma \mathbf{E}$. In such a
medium, propagation of an electromagnetic wave is determined by a
wave equation in the form
\begin {equation}
-\vec{\nabla}^2 \vec{E} + \mu_0 \sigma \frac{\partial
 \vec{E}}{\partial t} + \frac{1}{c^2}\frac{\partial^2  \vec{E}
}{\partial^2 t} = -\mu_0 \frac{\partial^2  \vec{P}}{\partial^2 t}.
\end {equation}
Employing the slowly varying amplitude approximation, which is
described by conditions,
\begin {eqnarray}
\frac{\partial \varepsilon}{\partial \mathrm{t}} \ll \omega
\varepsilon &,& \,
\frac{\partial \varepsilon}{\partial \mathrm{z}} \ll \mathrm{k} \varepsilon \\
\frac{\partial \mathcal{P}}{\partial \mathrm{t}} \ll \omega
\mathcal{P} &,&\, \frac{\partial \mathcal{P}}{\partial \mathrm{z}}
\ll  \mathrm{k} \mathcal{P}.
\end {eqnarray}
wave equation can be reduced to
\begin{equation} \label{D2}
\frac{\partial E}{\partial z} + \alpha E + \frac{1}{V_G}
\frac{\partial E}{\partial t} + i \,b_{2} \frac{\partial^2
E}{\partial^2 t}  = 0.
\end{equation}
Here, $E$ loss term $\alpha$, group velocity $V_G$, and dispersion
coefficient $b_{2}$ are given by \cite{Harris}
\begin{eqnarray}
\alpha &=& -\frac{i \pi}{\lambda} \chi(\omega_{0}),\quad
\frac{1}{V_G} = \frac{1}{c} - \frac{\pi}{\lambda} \frac{\partial
\chi}{\partial \omega}\\
b_{2}&=& \frac{\pi}{\lambda} \frac{1}{2}
 \left[\frac{\partial^{2}\chi}{\partial^{2}\omega}|_{\omega_{0}}\right].
\end{eqnarray}
\section{Numerical results and discussion}
\begin{figure}[htbp]
\begin{minipage}{18pc}
\includegraphics[width=18pc]{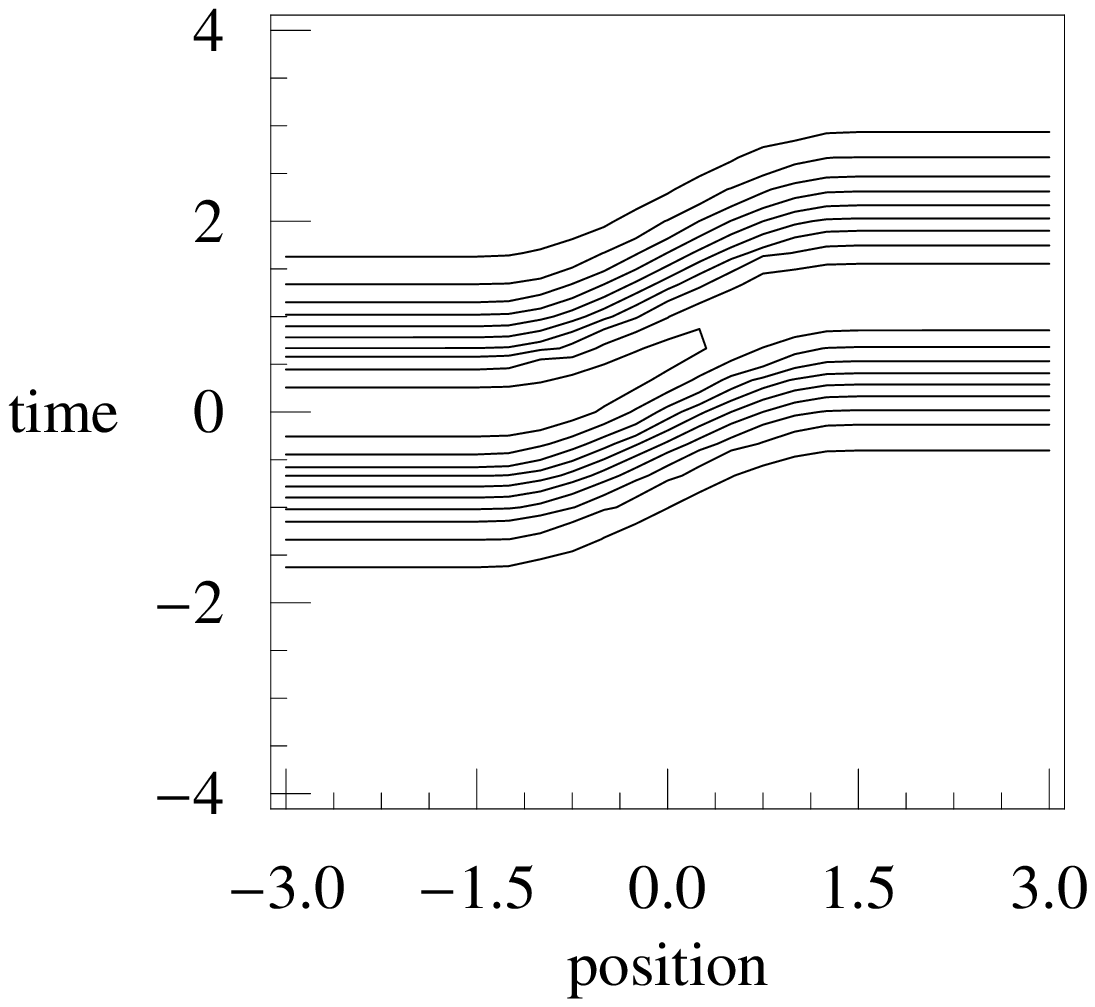}
\caption{\label{fig1}Contour plot of the propagation of an EIT
pulse of width $\tau=10^{-6}$s through a $^{23}$Na BEC. Time and
position are in dimensionless units as explained in the text.}
\end{minipage}\hspace{2pc}%
\begin{minipage}{18pc}
\includegraphics[width=18pc]{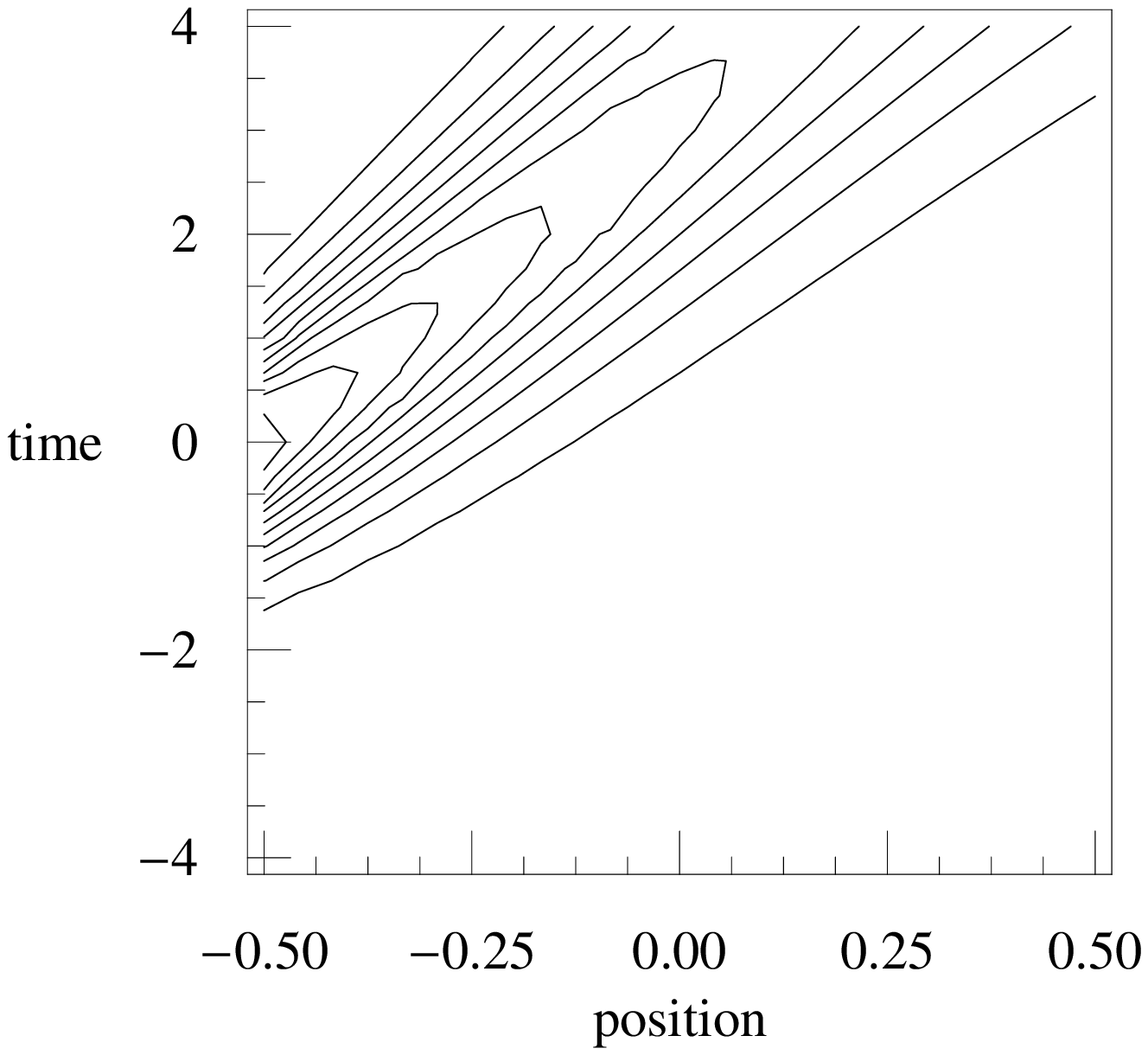}
\caption{\label{fig2}Same with Fig.\ref{fig1} but for
$\tau=10^{-7}$s.}
\end{minipage}
\end{figure}

Let us first compare the relative strength of the last two terms
of the wave equation. The third term describes the group velocity
dispersion while the last (fourth) one is for the second order
dispersion. We have analytically calculated the second order
dispersion coefficient $b_2$. Substituting the numerical values
taken from experiment\cite{Hau}, where $a_{s}=2.75\,\mathrm{nm}$,
$\omega_{r}=2\pi\times69\,\mathrm{Hz}$,$\omega_{z}=2\pi\times21\,\mathrm{Hz}$,
we numerically evaluated it to be in the order of $b_2\sim
10^{-9}\,\mathrm{s}^2/\mathrm{m}$ about the center of a $^{23}$Na
condensate with the total number of atoms $N=8.3\times10^6$. For
some other set of experimentally accessible parameters we have
found $b_2\sim 10^{-8}\,\mathrm{s}^2/\mathrm{m}$. We have also
checked third order dispersion coefficient and have found it $\sim
10^{-15}\,\mathrm{s}^3/\mathrm{m}$, which is insignificant. Taking
the ratio of the third term to the fourth term, we find that the
condition for dispersion to be significant is $(\tau/V_G)\times
10^9 \ll 1$, where $\tau$ denotes the temporal pulse width .
Subluminal light speeds achieved using $^{23}$Na BECs are in the
order of $1-100\mathrm{(m/s)}$. This yields $\tau \ll 10^{-8}$.
This explains clearly laser pulses used in present slow-light
experiments in condensates do not suffer from dispersion. They are
however not the optimal choice for a higher capacity optical
memory. One could use a ten times shorter pulse so that $\tau\sim
10^{-7}$. Such a pulse would be influenced by little dispersion in
the condensate and would be an optimal choice to use in optical
memories. We present our numerical simulation results in
Figs.\ref{fig1}-\ref{fig2} for the propagation of pulses of widths
$10^{-6}$s and $10^{-7}$s, respectively. In the figures we use
dimensionless scaled position ($z$) and time variables. Position
is scaled by the condensate radius while time is scaled by the
pulse width. Diameter of the BEC is $\sim50\,\mathrm{\mu m}$.
Effect of second order dispersion is a small broadening in the
temporal pulse width as demonstrated in Fig.\ref{fig2}, where,
behavior of the pulse about the center of the condensate, where
the dispersive effects are most influential, is shown.

Detailed investigation of behavior of even shorter pulses require
more intensive numerical studies and will be published elsewhere.
Wave equation is solved via finite difference Crank-Nicholson
space marching scheme. The Crank-Nicholson scheme is less stable
but more accurate than the fully implicit method; it takes the
average between the implicit and the explicit
schemes\cite{Garcia}. We use forward difference scheme for the
position in order to calculate the next step and we apply central
difference scheme for the time. We discretized the wave equation
and a set of linear equations are solved at each step to find
$E^{i+1}$ where $i$ represents the position z. At each step we
find $E^{i+1}$ but this method is too slow. Accordingly, we use
Thomas algorithm \cite{Garcia} which is equivalent than gaussian
elimination method for tri-diagonal matrices.

\section{Conclusion}

Ultra-slow light, which is achieved via EIT in BECs, has an
effectively long optical path lengths within the condensates.
Therefore dispersive effects, in particular temporal broadening of
the pulse, can be seen within the short (micrometer) length of
BEC. Normally, one would need much longer distances to be covered
with the pulse propagating in other more usual medium such as an
optical fiber. Though dispersion has no effect on the slowing down
the pulse, slow group velocity makes the dispersion visible in
such a short distance.

We examined propagation of short laser pulses in a Bose-Einstein
condensate taking into account the dispersive effects under the
conditions for electromagnetically induced transparency. Our
calculation of the high order dispersion coefficients using
typical experimental parameters of slow-light schemes show that
dispersive effects start to become influential for pulses whose
widths are about $10^{-7}$s. Present experiments uses microsecond
pulses which are below that dispersion limit.

\acknowledgements We would like to thank N. Postac\i{}o\~{g}lu and
A. Sennaro\~{g}lu for their useful discussions. This work was
supported by \.Istanbul Technical University Foundation (ITU BAP).
O.E.M acknowledges support from TUBA-GEB\.IP Award.

\end{document}